\documentclass[aps,pra,showpacs,amsmath,amssymb,amsfonts,lengthcheck,twocolumn,longbibliography]{revtex4-1}
\usepackage{dcolumn}    
\usepackage{bm} 
\usepackage{graphicx}
\usepackage{amsmath}    
\usepackage{latexsym}
\usepackage{amsfonts}   
\usepackage{amssymb}
\usepackage{array}      
\usepackage{epsfig}
\usepackage{txfonts}
\usepackage{xcolor}
\usepackage[colorlinks=true,linkcolor=blue,urlcolor=blue,citecolor=blue,pdfusetitle]{hyperref}
\usepackage{hyperref}
\usepackage{ulem}

\newcommand{\ket}[1]{\left|#1\right\rangle}

\newcommand{\braket}[2]{\langle#1|#2\rangle}
\newcommand{\ketbra}[2]{|#1\rangle\langle#2|}

\begin{document}
\title{Fast and robust magnon transport in a spin chain}
\author{Anthony Kiely}
\author{Steve Campbell}
\affiliation{School of Physics, University College Dublin, Belfield Dublin 4, Ireland}

\begin{abstract}
A protocol for fast and robust magnon transport in a one-dimensional spin chain is devised. Employing an approximate mapping between the chain and a single harmonically trapped particle, we exploit the known analytic control protocols for the latter and adopt them to achieve fast, high-fidelity transport in the chain. We compare the performance with finite time adiabatic protocols, establishing that the designed scheme allows for significantly faster and more stable transport. Furthermore, we show that a sharp transition exists between regions in which the protocol is effective and when it breaks down, giving rise to a heuristic speed limit for the process.

\end{abstract}
\date{\today}
\maketitle

\section{Introduction}
Achieving coherent transfer of information between distant regions of a quantum device is a critical task in the development of reliable quantum communication~\cite{DiVincenzo2000}. The seminal work of Bose established that low-dimensional spin chains can serve as excellent quantum communication channels where the information is encoded into single excitations within the chain~\cite{Bose2003}. These excitations travel as a spin wave, or quasi-particle, known as a magnon, whose utility in various quantum information processing tasks has been explored in recent works~\cite{Khitun2001,Song2005,Chen2017}.

In the case of a single excitation localised to one site at the end of an open spin chain, perfect transfer can be achieved sole through the free evolution of the chain. Without any further engineering, and neglecting any spoiling effects, such perfect state transfer only occurs for very specific times. However, subsequent works have developed this concept in a variety of ways, such as, by choosing non-uniform inter-spin couplings~\cite{Christandl2004, Tony1, Tony2}, employing dynamical control to the end spins achieving effective swap operations~\cite{Burgath2010}, exploiting topological systems~\cite{Graham2020, Impens2020, Greek2020}, and using adiabatic evolutions~\cite{Korzekwa2014}. While these schemes are well suited for the transport of a single localised excitation, a freely evolving magnon will tend to spread throughout the chain due to dispersion of the wave-packet. Therefore, alternative approaches, which are able to transport an arbitrarily sized magnon wave-packet, offer a wider range of applicability and are therefore desirable. 

One such alternative dynamic approach is to use a specific spatio-temporal profile of the magnetic field where the control scheme is designed using the adiabatic theorem in combination with analogies with classical mechanical systems~\cite{Gong2008} or light guided in waveguides~\cite{Makin2012,Ahmed2015,Ahmed2017,Gruszecki2018}. The trapping potential in these cases requires only ``semi-local" control~\cite{Ahmed2015} since the magnetic gate resolution can be relatively large compared with the inter-spin spacing. Nevertheless, these techniques tend to require long operation times opening the possibility to the system deteriorating due to decoherence. To address this issue schemes to achieve high fidelity control on non-adiabatic timescales have been developed, known as shortcuts to adiabaticity~\cite{STAreview}. These techniques are analytical methods to control quantum, classical and even biological systems~\cite{Iram2020} in fast total operation times. They are extremely effective for several ideal systems, such as, harmonic oscillators~\cite{Torrontegui2011} and few level systems~\cite{Ruschhaupt2012, Benseny2017, Kiely2016}. However, the methods are more challenging to apply to complex, many-body systems as the protocols often require exactly solving the model. To address this issue, several alternative approaches have been followed, for example by approximating counterdiabatic terms with only allowed operations~\cite{Opatrny2014,Saberi2014,Campbell2015,Sels2017} or improving existing schemes for idealisations using optimal control~\cite{Mortensen2018, Abah2019} or perturbation theory~\cite{Whitty2020}.
 
In this work, we follow a similar approach and develop a robust and stable scheme to reliably transport a single excitation wave-packet (magnon)~\cite{Kranendonk1958, Parkinson2010, Stancil2008} on non-adiabatic timescales. By exploiting an approximate mapping between the spin-chain and a harmonically trapped single particle, we employ the known analytic techniques from shortcuts to adiabaticity to the transport of the magnon. The resulting operation timescales are significantly better than previous adiabatic schemes, and furthermore, the protocol is inherently stable. In addition to the clear practical advantages, our approach also demonstrates that the insight gained from analytical control schemes~\cite{Sels2017, Puebla2020} for simple systems are invaluable in developing useful control protocols in complex many-body systems.

After reviewing the system Hamiltonian for the Heisenberg spin chain in Sec.~\ref{sys}, in Sec.~\ref{map} we outline how magnon transport in this case can be mapped to the transport of a single particle. In Sec.~\ref{sta} we design the variation of the magnetic field for near-perfect transport using the methods of shortcuts to adiabaticity~\cite{STAreview}. The effectiveness of these new schemes compared to previous adiabatic techniques are demonstrated in Sec.~\ref{numerics} where the maximal transport velocities and the effect of disorder are also investigated. Finally, in Sec.~\ref{con} we summarise our results and outline possible future work. 

\section{Preliminaries}
\subsection{Model \label{sys}}
The time-dependent Hamiltonian for $N$ spins which we consider is given by
\begin{equation}
H(t)= - \frac{J}{2} \sum_{n=1}^{N-1} \vec{\sigma}_{n}\cdot \vec{\sigma}_{n+1}+ \sum_{n=1}^{N} B_n(t) \sigma_n^z, \label{heis}
\end{equation}
where $\vec{\sigma}_n=(\sigma_n^x,\sigma_n^y,\sigma_n^z)$, $J$ is the interaction strength due to dipole-dipole or exchange interactions, and $B_n(t)$ is the local magnetic field strength at site $n$ where the gyromagnetic ratio has been absorbed into this definition. We will use the notation that the spin down and up states are given by $\ket{0}=\ket{\downarrow}$ and $\ket{1}=\ket{\uparrow}$. Hamiltonian \eqref{heis} preserves the total excitation number, i.e., $\left[H(t), \sum_{n=1}^N \sigma_n^z \right]\!=\!0 $ which allows us to define a basis for the single excitation subspace $\ket{\phi_n}=\otimes_{m=1}^N \ket{\delta_{m,n}}$. Hamiltonian \eqref{heis} in this basis and subspace then becomes~\cite{Murphy2010}
\begin{equation}
H_s(t)=H_0+H_1(t), \label{mapped}
\end{equation}
where 
\begin{eqnarray}
H_0&=&-2J \sum_{n=1}^N  \ketbra{\phi_n}{\phi_n} + J \Big[ \ketbra{\phi_1}{\phi_1}+\nonumber \\ 
			&& \ketbra{\phi_N}{\phi_N} +  \sum_{n=1}^{N-1} \left( \ketbra{\phi_n}{\phi_{n+1}} + \ketbra{\phi_{n+1}}{\phi_n} \right) \Big],  \\ 
H_1(t)&=&\sum_{n=1}^N B_n(t) \ketbra{\phi_n}{\phi_n}. 
\end{eqnarray}
Working with the Hamiltonian in this subspace greatly improves numerical efficiency since it is dimension $N$ in contrast with the typical exponential scaling $2^N$.

Our goal is to quickly transfer a single excitation wavepacket a given distance across the spin-chain. Previous works that considered adiabatic transport schemes have explored different spatial profiles for the applied field, $B_n(t)$, including parabolic~\cite{Gong2008}, P{\"o}schl-Teller~\cite{Makin2012,Ahmed2015} and square well~\cite{Ahmed2015}. In what follows we will assume the magnetic field has a time-dependent parabolic spatial profile, however, we remark that our scheme can be readily be applied to other potentials which are locally harmonic. In order to develop our control protocol we begin by mapping Eq.~\eqref{mapped} to a single trapped particle for which analytic techniques for perfect transport are known.

\subsection{Continuum Limit and Approximate Mapping \label{map}}
The Hamiltonian for a single particle in a potential is given by
\begin{eqnarray}
H_c(t) &=& \frac{p^2}{2m}+V(x,t). \label{singlep}
\end{eqnarray}
We consider the resulting Hamiltonian when the $x$ coordinate is discretized into points $x_m$ with uniform spacing $\Delta x\!=\! x_{m+1}-x_m$ where $m=1,2,...,N$. The kinetic term becomes 
\begin{equation}
\frac{\hbar^2}{2 m \left(\Delta x\right)^2} \left[ \sum_{n=1}^N  2\ketbra{x_n}{x_n} - \sum_{n=1}^{N-1} \left( \ketbra{x_n}{x_{n+1}} + \ketbra{x_{n+1}}{x_n} \right)\right], \label{kinetic}
\end{equation}
where $\ket{x_n}$ is an eigenstate of the position operator with eigenvalue $x_n$. Since we are working in position space, the potential is diagonal, i.e. $V(x,t)\!=\!\sum_{n=1}^N V(x_n,t) \ketbra{x_n}{x_n}$. Note that we are assuming boundary conditions such that the wave-function is exactly zero outside the discretised space i.e. ``hard wall" boundary conditions. The continuum limit $\Delta x \rightarrow 0$ and $N \rightarrow \infty$ with $N \Delta x$ a constant recovers the Hamiltonian for a single particle, Eq.~\eqref{singlep}.

We can now make a correspondence between the discretized version of the single particle Hamiltonian $H_c$ and that of the Heisenberg spin chain in the single excitation subspace Eq.~\eqref{mapped}~\cite{Makin2012}, which is the starting point of our control protocol. The single particle potential clearly plays the role of the local magnetic field $V(x_n,t)\!\sim\!B(x_n,t)=B_n(t)$ where $x_n\!=\!(n-1)\Delta x$ is the position of site $n$ with inter-spin spacing $\Delta x$. Aside from some edge effects (arising from the open boundary conditions), the kinetic energy term, Eq. \eqref{kinetic}, is analogous to the nearest neighbour coupling in Eq.~\eqref{mapped}, which allows for the connection $- \frac{\hbar^2}{2 m \left(\Delta x\right)^2}\!\sim\!J$.

Using this mapping for a harmonic trapping potential
\begin{equation}
    V(x_n,t)=\frac{1}{2} m \omega(t)^2 \left[x_n-X_0(t)\right]^2, \label{harmonic_pot}
\end{equation}
the time-dependent term of the spin system Hamiltonian in the single excitation subspace becomes
\begin{equation}
\label{H1mapped}
H_1(t)=-\frac{\hbar^2 \omega(t)^2}{4 J}\sum_{n=1}^N \left[\frac{x_n-X_0(t)}{\Delta x}\right]^2 \ketbra{\phi_n}{\phi_n}.
\end{equation}
Physical control of the system via the magnetic field $B_n(t)$ is now parameterised by two control functions $\omega(t)$ and $X_0(t)$. In the next subsection, we will outline how they are determined.

For a harmonic trap, the wave-function width is $\sigma\!=\!\sqrt{\frac{\hbar}{|m| \omega}}$. One would expect this mapping to be very accurate when $\sigma \!\gg\! \Delta x$ which reduces to $\omega\!\ll\! 2 J/\hbar$. Heuristically, this implies that excessively tight trapping is not possible as a very narrow wave-packet is not be resolvable on the finite grid. As such, the mapping is not readily applicable to the transportation of a single excitation localised to a single site.

Finally a few important remarks about the mapping. Firstly there is a discrepancy between the Hamiltonians at the edges. To circumvent this, in what follows we will assume that our wave-packet is initially prepared within the bulk of a larger spin-chain register and therefore is kept sufficiently far from the edges. While this ensures that the boundary conditions do not play a significant role, we continue to use open boundary conditions to remain consistent with previous results on magnon transport. Secondly, a positive value of $J$ implies the magnon has a negative effective mass. Treating everything consistently results in an inverted trapping potential. While this experimentally challenging for the case of a trapped particle, this does not present an issue in the spin chain scenario.

\subsection{Shortcuts to Adiabaticity for a Harmonically Trapped Particle \label{sta}}
Here we will briefly review invariant based inverse engineering for a single particle in a harmonic trapping potential being transported~\cite{Torrontegui2011}, which by virtue of the mapping outlined in Sec.~\ref{map}, will be used to design the spatio-temporal profile of the magnetic field in Eq.~\eqref{H1mapped} to achieve high fidelity magnon transport.

The single particle Hamiltonian $H_c$ with a harmonic potential (see Eq. \eqref{harmonic_pot}), has a Lewis-Riesenfeld invariant given by~\cite{Torrontegui2011}
\begin{equation}
I=\frac{1}{2m}\left[\rho (p-m \dot{X}_c)- m \dot{\rho}(x-X_c)\right]^2+\frac{1}{2}m \omega_0^2 \left(\frac{x-X_c}{\rho}\right)^2.
\end{equation}
Using the property that $I$ does not vary in time implies that $\frac{d I }{d t}\!=\!\frac{\partial I}{\partial t}+\frac{1}{i \hbar} \left[I,H\right]=0$ which leads to two auxiliary equations
\begin{eqnarray}
&\ddot{\rho}+\omega(t)^2 \rho = \frac{\omega_0^2}{\rho^3}, \label{aux1} \\
&\ddot{X}_c+\omega(t)^2 (X_c-X_0) = 0. \label{aux2}
\end{eqnarray}
The solution of the corresponding Schr{\"o}dinger equation in this case is $\ket{\psi(t)}\!=\!\sum_{n=0}^\infty c_n e^{i \alpha_n(t)} \ket{\psi_n(t)}$ where the $c_n$ are time independent complex coefficients, the modes $\ket{\psi_n}$ are eigenstates of the Lewis-Riesenfeld invariant $I(t) \ket{\psi_n(t)}\!=\!\lambda_n \ket{\psi_n(t)}$ and the Lewis-Riesenfeld phases are given by $\alpha_n(t)\!=\!-\frac{1}{\hbar} \int_0^t ds \left[ \frac{\lambda_n}{\rho^2}+\frac{m(\dot{X}_c \, \rho-X_c \, \dot{\rho})^2}{2 \rho}\right]$, while the modes in coordinate space are given by 
\begin{equation}
\braket{x}{\psi_n(t)}= \exp \left\{\frac{i m}{\hbar} \left[ \dot{\rho} x^2/2 \rho +(\dot{X}_c \, \rho-X_c \, \dot{\rho})/\rho \right] \right\} \rho^{-1/2} f\left(\frac{x-X_c}{\rho}\right),
\end{equation}
where $f(q)$ is an eigenstate of a quantum harmonic oscillator with potential $\frac{1}{2} m \omega_0 q^2$ and eigenvalues $\lambda_n=(n+1/2) \hbar \omega_0$.

Let us assume that the harmonic trap is moved from $X_0(0)\!=\!x_A$ to $X_0(t_f)\!=\!x_B$. To ensure perfect state transfer between the associated ground state wavepackets, the following boundary conditions must be enforced 
\begin{equation}
X_c(0)= x_A,~~~X_c(t_f)= x_B,~~~\frac{d^n X_c(t')}{d t^n}=0 \label{Xc_bc}
\end{equation}
for $n=1,2$ and $t'\!=\!0,t_f$. To allow for extra control, the trap frequency is also changed from $\omega(0)\!=\!\omega_0$ to $\omega(t_f)=\omega_f$ which requires the additional boundary conditions
\begin{equation}
\rho(0)=1, ~~~\rho(t_f)=\gamma=\sqrt{\frac{\omega_0}{\omega_f}},~~~\frac{d^n \rho(t')}{d t^n}=0, \label{rho_bc}
\end{equation}
for $n=1,2$ and $t'\!=\!0,t_f$. The required trap trajectory and frequency can be found by inverting the auxiliary equations
\begin{eqnarray}
\omega(t)^2 &=& \frac{1}{\rho} \left[ \frac{\omega_0^2}{\rho^3}-\ddot{\rho} \right], \label{control1} \\
X_0(t) &=& \frac{\ddot{X}_c}{\omega^2(t)}+X_c. \label{control2}
\end{eqnarray} 

Now by choosing auxiliary functions, which fulfil the boundary conditions Eqns. \eqref{Xc_bc} and \eqref{rho_bc}, one can find the variation of the control parameters $\omega(t)$ and $X_0(t)$ to ensure perfect state transfer in the single particle setting. Note that anharmonicities will introduce a lower bound on the operation time $t_f$~\cite{Exp2020}. By virtue of the correspondence between the single particle Hamiltonian $H_c$ and the single excitation subspace Hamiltonian $H_s$ (see Sec. \ref{map}), intuitively we expect that these simple analytical protocols will continue to work  effectively for transport in the chain. In the next section we demonstrate this numerically for a range of operation times.

It is worth emphasising that our approach has the advantage of being principally analytical, and therefore does not necessarily require numerical optimization. Nevertheless, as the method relies on the approximate validity of the mapping,  it invites the possibility to use the analytic solutions as initial seeds for numerical optimal control. Indeed, similar ``hybrid" schemes have already proven to be highly effective for achieving control in complex many-body systems~\cite{Campbell2015, Sels2017, Saberi2014}.
\begin{figure*}[t]
\begin{center}
\includegraphics[width=0.99\linewidth]{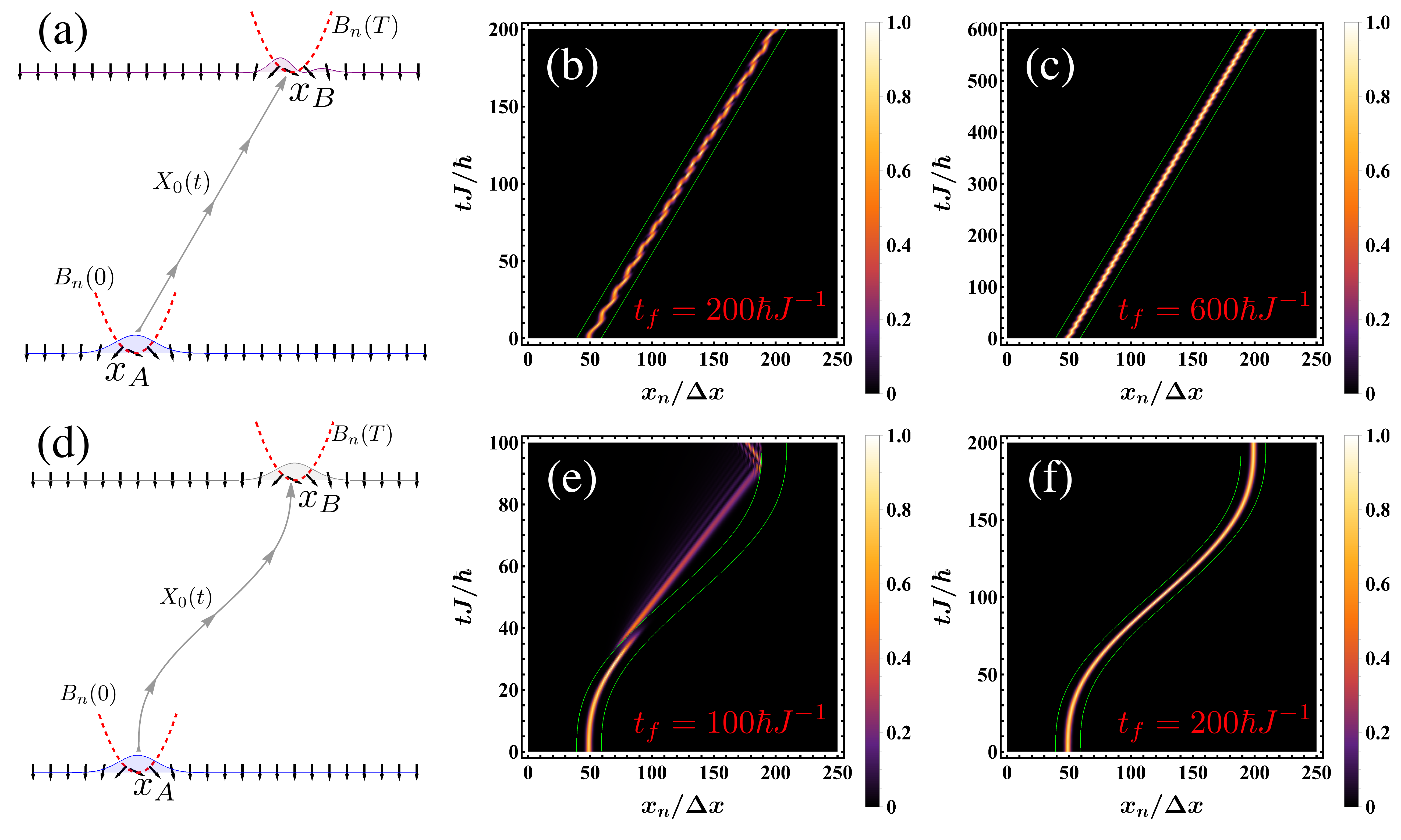} 
\end{center}
\caption{(a-c) Adiabatic scheme, Eq~\eqref{adiabtraj}. (a) Schematic of the spin-chain (black arrows) and the magnitude of magnetic field profile (dashed red line). An adiabatic protocol (grey arrows) transports the initial state at $x_A$ (blue shaded region) and to an imperfect final state (purple shaded region) when $t_f$ is small. (b) and (c) Variation in local magnetisation, $\langle \sigma_z^n \rangle$,  against time $t$ and spin-site $x_n$. In panel (b) $t_f\!=\!200 \, \hbar J^{-1}$ which is significantly faster than adiabatic timescales and results in very poor magnon transport ($F\!\sim\!0.3$) while in (c) $t_f\!=\!600 \, \hbar J^{-1}$ which is approaching adiabatic timescales.
(d-f) Nonadiabatic scheme Eq.~\eqref{statraj}. (d) Similar schematic, however now showing the nonadiabatic scheme (grey arrows) and the resulting perfect final state (grey shaded region). (e) and (f) Variation in local magnetisation, $\langle \sigma_z^n \rangle$  against time $t$ and spin-site $x_n$ for (b) $t_f\!=\!100 \hbar J^{-1}$ where our protocol fails due to the transport occurring too fast and (c) $t_f=200 \hbar J^{-1}$ which ensures near perfect magnon transport. In all panels the boundary of trapping potential is shown in green. Other parameters used are $N=251$, $\omega_0\!=\!0.5 J/\hbar$, $x_A \!=\! 50 \, \Delta x$ and $d\!=\!150 \, \Delta x$.}
\label{fig1}
\end{figure*}

\section{Fast and Robust Magnon Transport \label{numerics}}

In this section we will compare our approximate analytical nonadiabatic control scheme (see Eqns. \eqref{control1} and \eqref{control2}) with a previous adiabatic method~\cite{Gong2008} where a parabolic potential moving linearly in time was used. Due to the adiabatic theorem for very long operation times, the linear protocol was found to produce high fidelity state transfer of single excitation Gaussian wavepackets. We will demonstrate in the following that while the adiabatic protocols are ineffective for fast magnon transport, our approach of applying approximate analytic techniques to complex systems is fast, stable, and robust.

As initial and target states of the spin chain we consider 
\begin{equation}
\begin{aligned}
&\ket{\psi_I} = \frac{1}{\sqrt{\sum_n b^2_n(x_A)}}\sum_n b_n(x_A) \ket{\phi_n}, \\
&\ket{\psi_T} = \frac{1}{\sqrt{\sum_n b^2_n(x_B)}}\sum_n b_n(x_B) \ket{\phi_n},
\end{aligned}
\end{equation}
where $b_n(u)\!=\!\text{Exp}\left[ {-(x_n-u)^2/ 2 \sigma^2}\right]$, which corresponds to a Gaussian wave-packet centred at $x_A$ and $x_B$, respectively~\cite{Song2005,Gong2008}. To avoid excessively high magnetic field strengths the potential is truncated such that it only acts $5 \sigma$ either side of $X_0(t)$, however in what follows the magnon is always sufficiently localised that this truncation does not have a significant effect on the dynamics. To quantify the efficacy of our protocol we will consider the instantaneous fidelity $F(t)=\left|\braket{\psi_T}{\psi(t)} \right|^2$. We will consider situations in which the magnon is transported a distance $d=x_B-x_A$, where $x_A$ and $x_B$ are chosen sufficiently far from the boundaries. We fix the size of the spin-chain to be $N\!=\!251$, however, we remark that qualitatively similar results are obtained for any other suitable value of $N$.

\subsection{Comparison of protocols}
For the adiabatic scheme~\cite{Gong2008}, we choose a linear ramp for the harmonic trap
\begin{equation}
X_0(t)=x_A+ s \, d, \label{adiabtraj}
\end{equation}
where $s\!=\!t/t_f$ and fix the trapping frequency to be a constant, $\omega(t)=\omega_0$, cfr. Fig. \ref{fig1}(a). If we consider long operation times, i.e. approaching adiabatic time scales, naturally this scheme works well. For example, in Fig. \ref{fig1}(c) a final fidelity of $F\!\sim\!0.998$ is achieved for a total operation time of $t_f=600 \hbar J^{-1}$. The magnon is being moved at a slow enough rate that the entire system remains essentially in equilibrium. However for much shorter times, $t_f\!=\!200 \hbar J^{-1}$, the final fidelity reduces to $F\!\sim\!0.3$. The dynamical behavior of the magnon wave-packet is shown in Fig. \ref{fig1}(b). Here we see that when moving at such high speeds the portion of the magnon that remains within the trapping potential oscillate significantly. Furthermore, not visible in the Fig. \ref{fig1}(b) is the small fragments of the excitation (approximately $0.1\%$) that get left behind as the potential moves though the chain.

For the non-adiabatic scheme we assume no change in the frequency $\omega(t)$ and therefore set $\rho\!=\!1$ in Eq. \eqref{control1}. In order to satisfy the remaining boundary conditions (Eq. \eqref{Xc_bc}) we choose the minimal polynomial ansatz for $X_c(t)\!=\!x_A+ d(6 s^5-15 s^4+10 s^3)$. From Eq. \eqref{control2}, we find that the trajectory in this case is given by
\begin{equation}
X_0(t)=x_A+d\left[6 s^5-15 s^4+10 s^3+\frac{60 s}{\omega_0^2 t_f^2}(1-3 s+2 s^2)\right]. \label{statraj}
\end{equation}
In Fig. \ref{fig1}(f) we again fix the operation time $t_f\!=\!200 \hbar J^{-1}$ to compare with the case where the adiabatic scheme failed to achieve high-fidelity transport, cfr. Fig. \ref{fig1}(b). Our scheme achieves a final fidelity of $F\!\sim\!0.998$, thus confirming it is an effective technique. However, while there are in principle no constraints on the time scales for perfect transport of a harmonically trapped atom following the recipe described in Sec.~\ref{sta}, the approximate nature of the mapping means this is not the case for the spin-chain. In Fig.~\ref{fig1}(e) we fix $t_f\!=\!100 \hbar J^{-1}$ where it is clear that the errors arising from the discrete nature of the system quickly leads to a breakdown in the effectiveness of the protocol. Infidelity resulting from the imperfect mapping to the continuum case could be mitigated following the enhanced shortcuts to adiabaticity approach~\cite{Whitty2020}.

\begin{figure}[t]
\begin{center}
\includegraphics[width=0.49\linewidth]{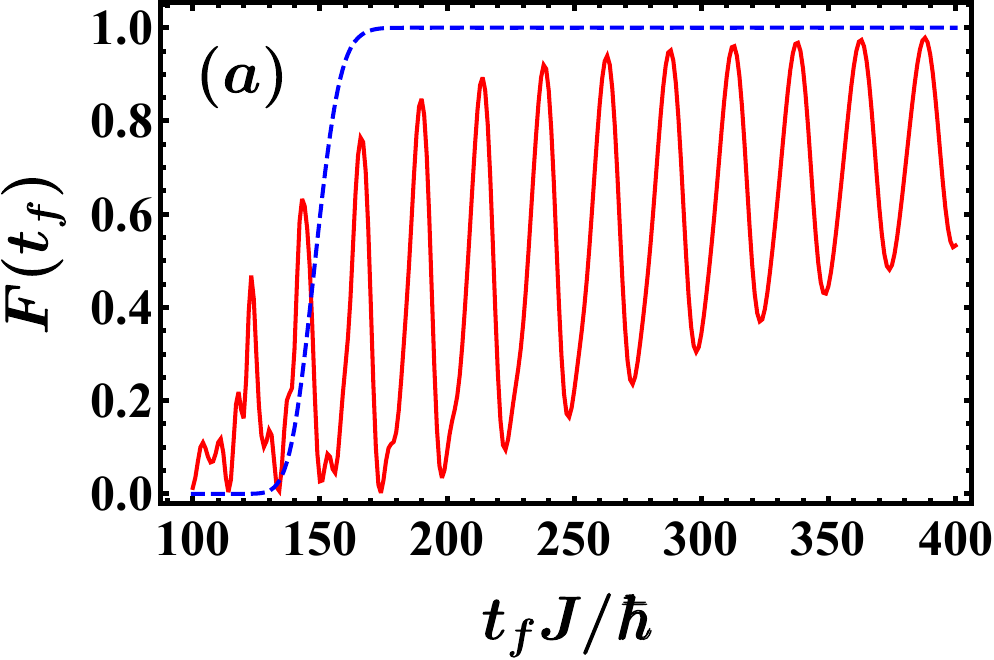} 
\includegraphics[width=0.49\linewidth]{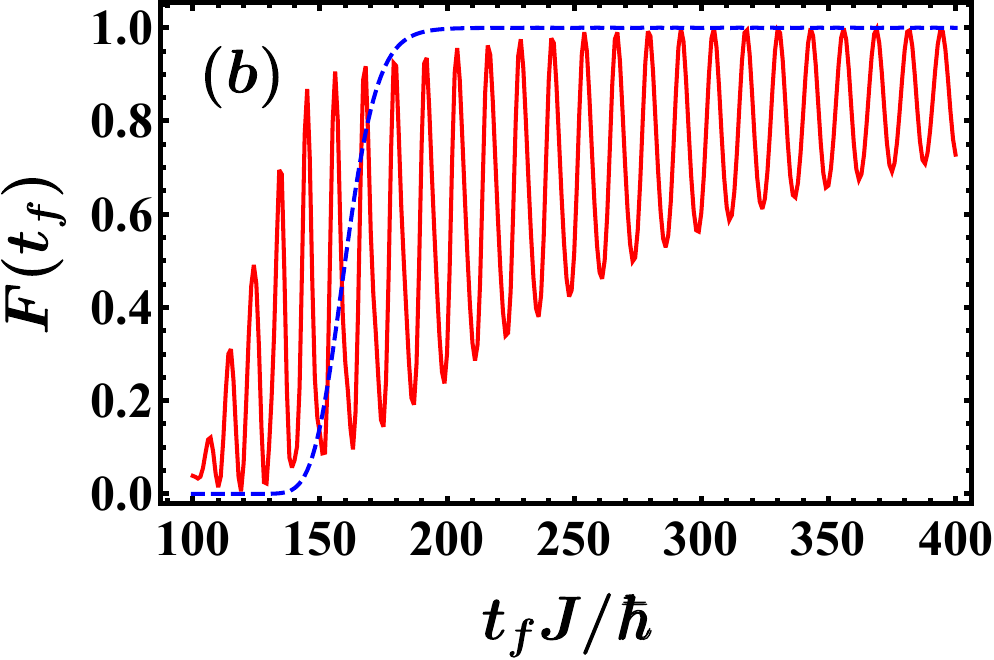}  \\
\includegraphics[width=0.49\linewidth]{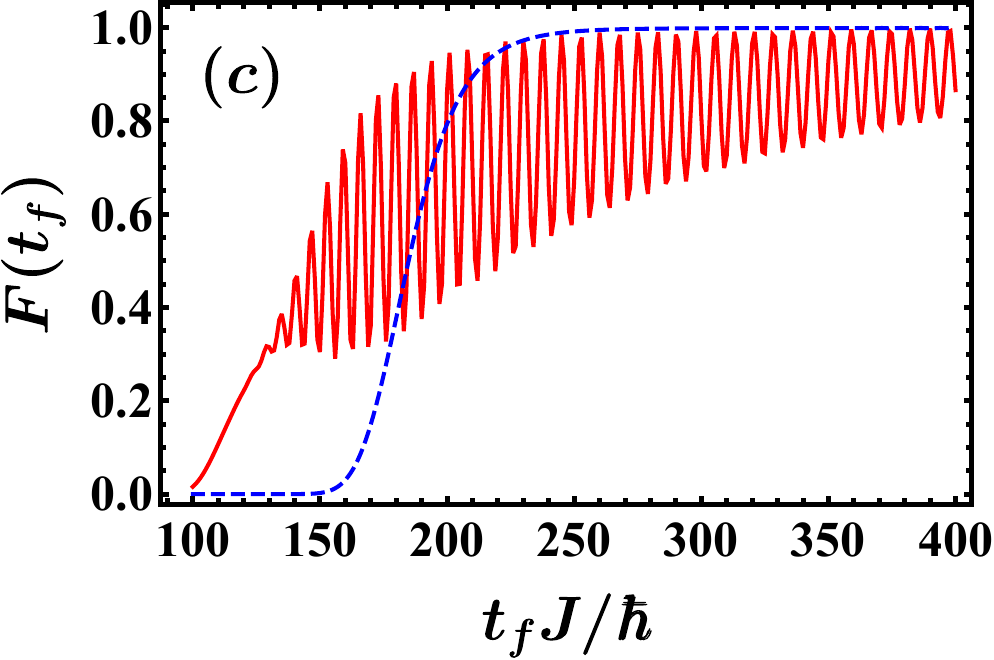}
\includegraphics[width=0.49\linewidth]{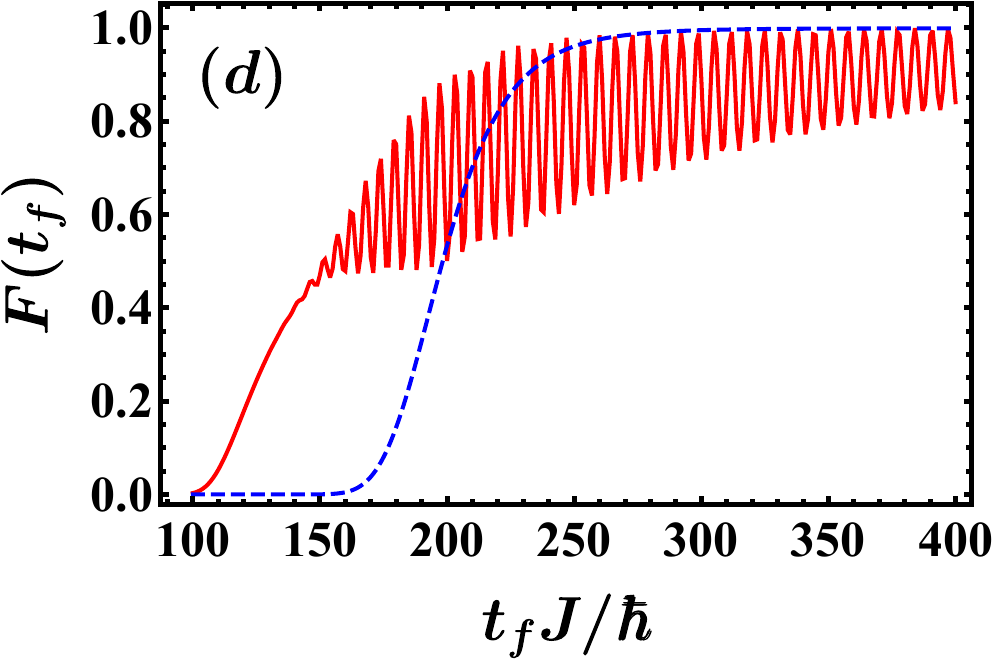}
\end{center}
\caption{Target state fidelity as a function of total operation time, $t_f$, for the adiabatic scheme, Eq.~\eqref{adiabtraj} (red, solid line) and the non-adiabatic protocol, Eq.~\eqref{statraj} (blue, dashed line). Each panel corresponds to a different value of the trapped frequency: (a) $\omega_0\!=\!0.25 J/\hbar$ (b) $\omega_0\!=\!0.5 J/\hbar$ (c) $\omega_0\!=\!0.85 J/\hbar$ (d) $\omega_0\!=\!1 J/\hbar$. In all panels we fix $x_A\! =\! 50 \, \Delta x$, $d\!=\!150 \, \Delta x$ and $N\!=\!251$.
\label{fig_sweep}}
\end{figure}

We can compare and contrast the effectiveness of both protocols by examining the final target state fidelity as a function of operation time, $t_f$, as shown in Fig.~\ref{fig_sweep} where we also consider the effect of different (fixed) trapping frequencies $\omega_0$. The red solid curves correspond to the adiabatic scheme, Eq.~\eqref{adiabtraj}, where a quite typical oscillatory behaviour is observed. As $t_f$ increases the magnitude of these oscillations decay indicating that we are approaching the adiabatic limit. In contrast, for the non-adiabatic scheme, Eq.~\eqref{statraj}, once the operation duration is large enough to ensure a high target state fidelity it remains so for larger values of $t_f$.

It is interesting to notice that for carefully chosen values of $t_f$ the adiabatic scheme can achieve high final fidelities on time-scales comparable to those where our non-adiabatic scheme is effective. These specific times are related to the concept of ``magic times''~\cite{Couvert_2008}. However, to achieve this requires very precisely tuning the operation time, an issue which is not present for the designed protocol. The robustness of our scheme will be further explored in Sec.~\ref{disorder}. In all panels we see that there is a range of values of $t_f$ during which fast transport is not possible, the precise values being dependent on the trapping frequency. This implies that there is a clear speed limit for the protocol. That such a minimal time should emerge is natural in light of the known bounds on the transmission of information in spin-chains. We will now investigate these limitation in detail in the next subsection.

\begin{figure}[t]
\begin{center}
\includegraphics[width=0.53\linewidth]{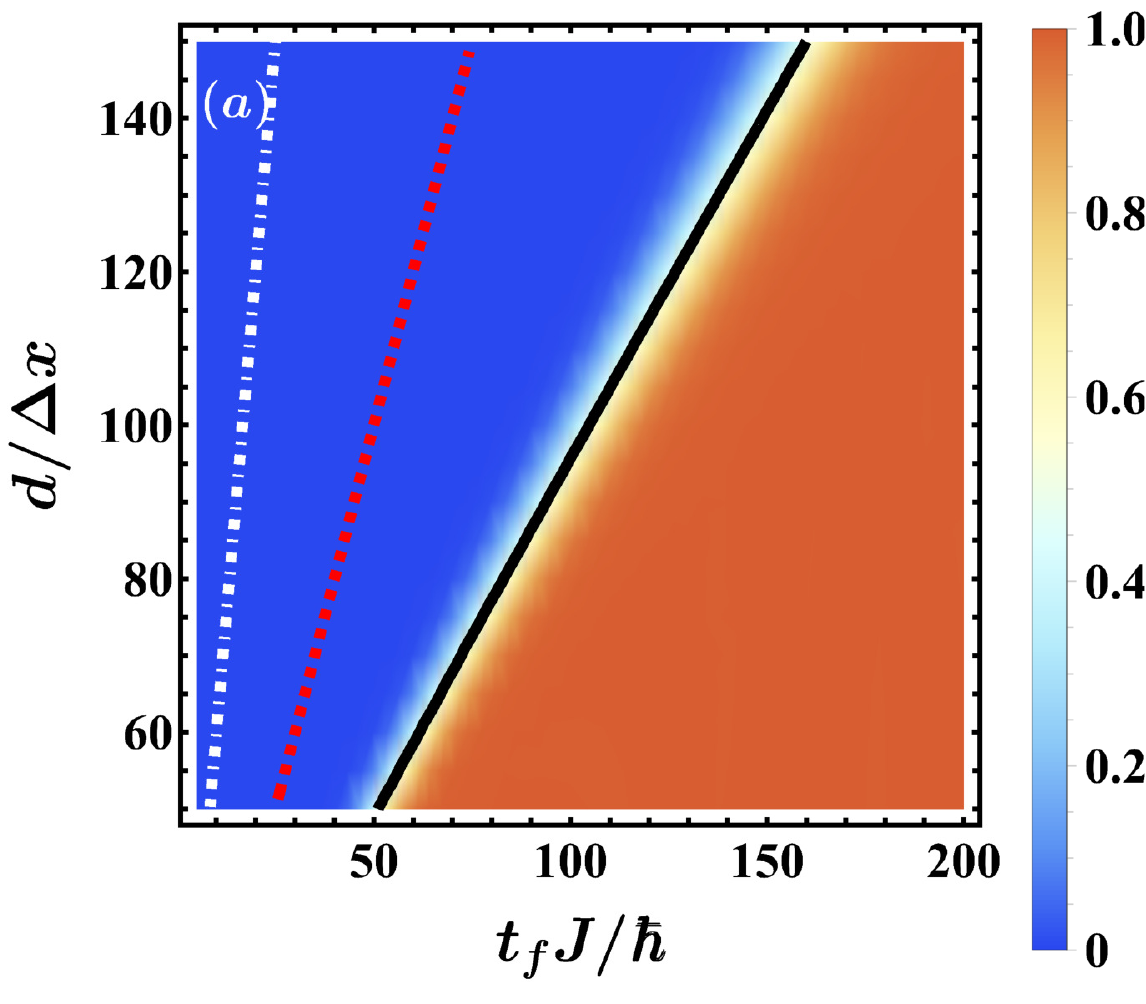} 
\includegraphics[width=0.45\linewidth]{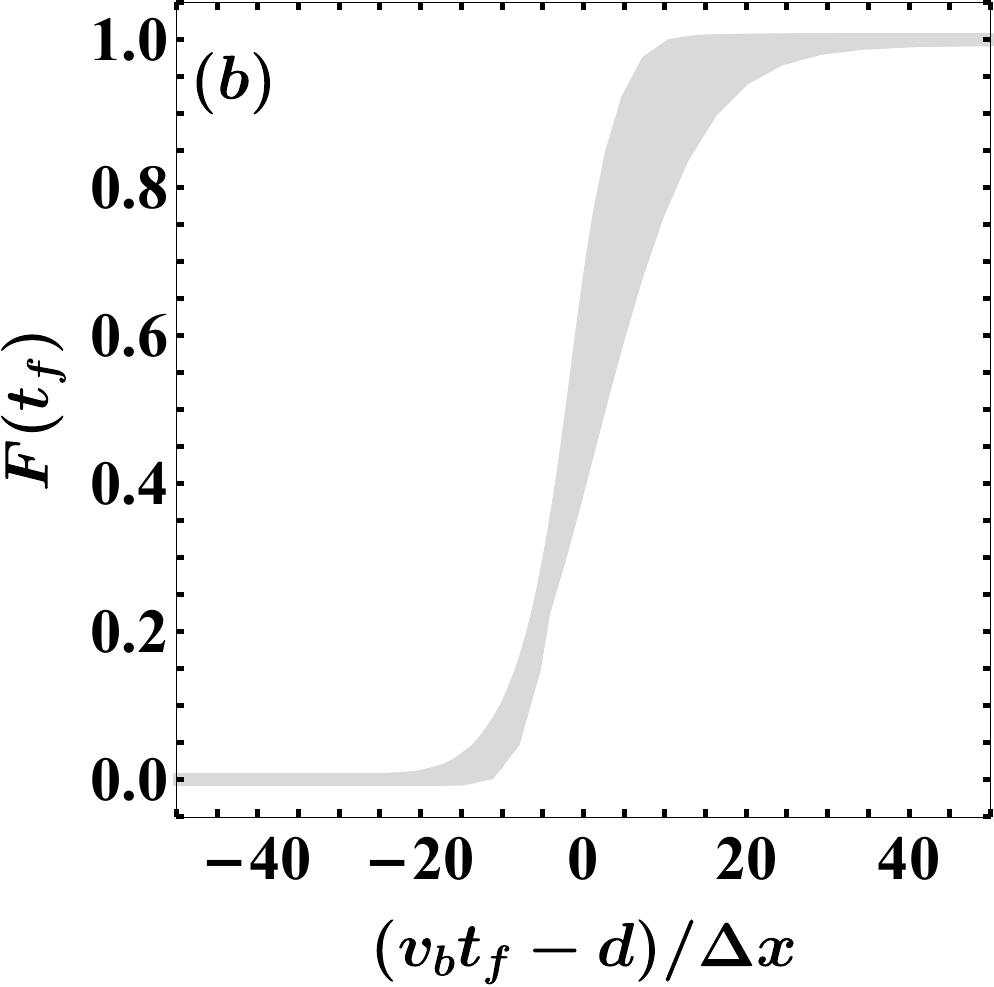} \\
\includegraphics[width=0.53\linewidth]{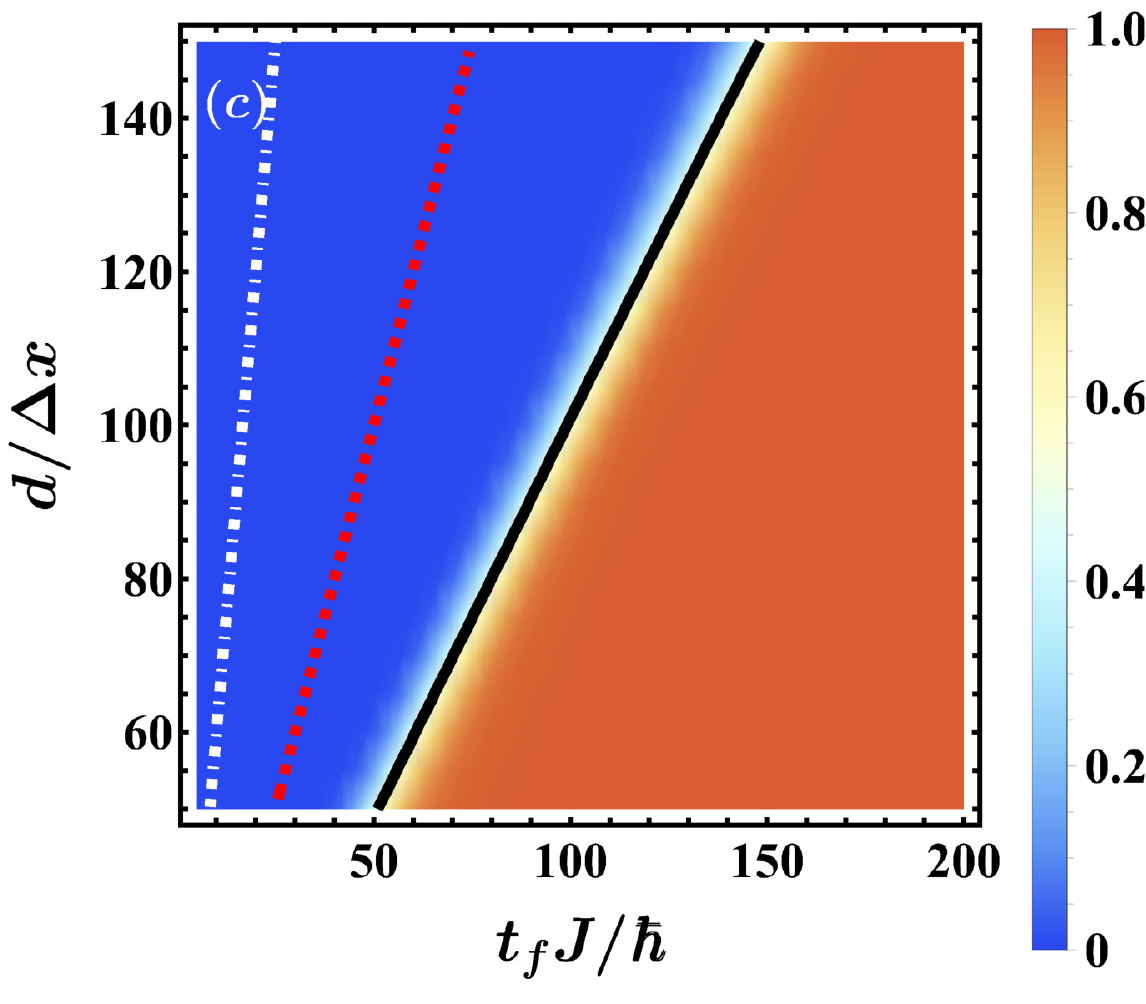} 
\includegraphics[width=0.45\linewidth]{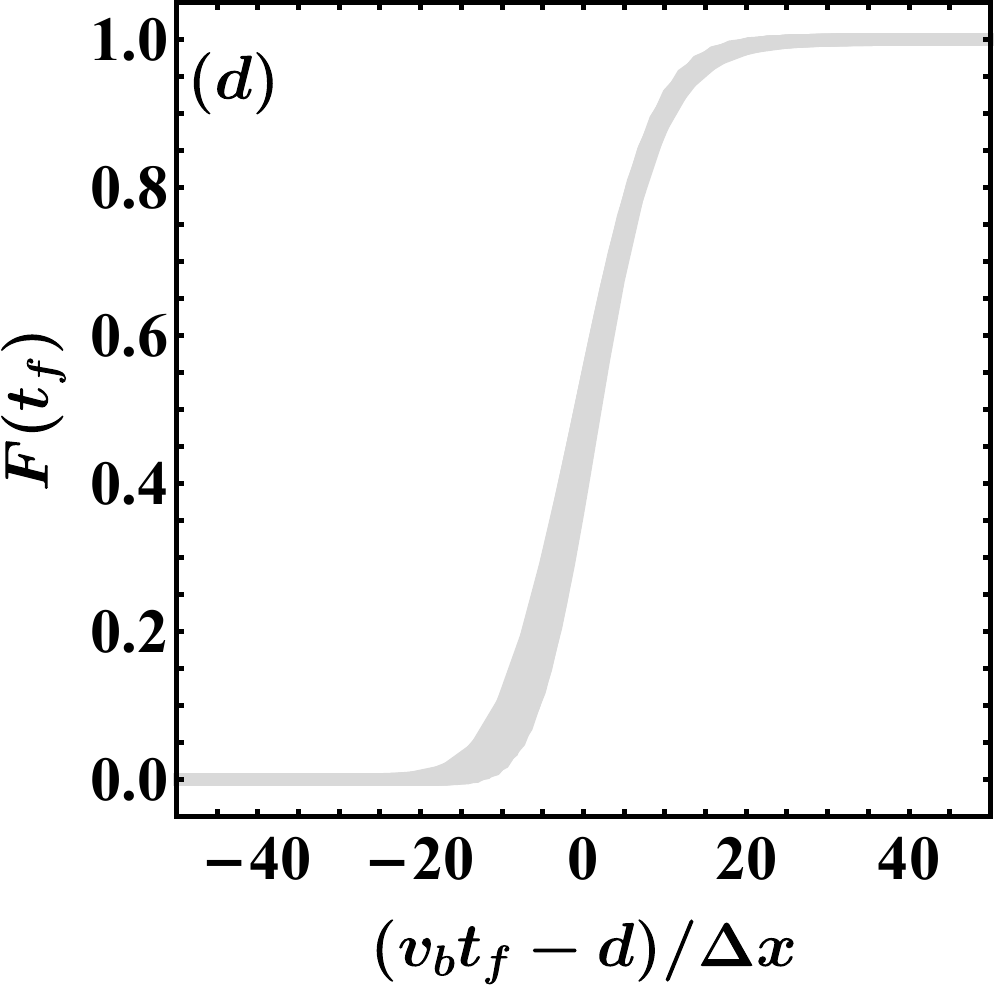} 
\end{center}
\caption{ Fidelity as a function of total operation time $t_f$ and distance $d$ for nonadiabatic scheme \eqref{statraj}. (a,c) Density plot with Lieb-Robinson velocity bound $6 \Delta x \, J /\hbar$ (white, dot-dashed line), maximum allowed group velocity  $2 \Delta x \, J /\hbar$ (red, dashed line) and numerically found limit velocity $v_b$ (black, solid line). (b,d) Transition between high and low fidelity regimes (for parameter ranges shown in (a,c)). Parameters: $x_A = 50 \, \Delta x$, $N=251$, (a,b) $\omega_0=0.5 J/\hbar$ and (c,d) $\omega_0=0.25 J/\hbar$.
\label{fig_sweepDT}}
\end{figure}
\subsection{Speed Limits for Magnon Transport}
Since the interactions in the Heisenberg spin chain Eq.~\eqref{heis} are local, the excitation can only propagate along the chain at a finite velocity. From the Lieb-Robinson bound~\cite{Lieb1972}, this velocity was found to be $6 \Delta x \, J/\hbar $~\cite{Epstein2017}. However in Ref.~\cite{Murphy2010} it was found that high fidelity numerically optimal protocols could not be found for velocities greater than $2 \Delta x \, J/\hbar $, which also corresponds to the maximal allowed group velocity in the system~\cite{Epstein2017, Ashab2012}, and thus represents an attainable speed limit for transport. Here we compare the maximal allowed velocity for our protocol which emerges numerically, with these known bounds.

In particular, in order to investigate further how well the designed non-adiabatic control protocols work when close to the temporal limit of the system Hamiltonian, we compute the fidelity as a function of both total operation time $t_f$ and distance $d$ shown in Fig. \ref{fig_sweepDT}. We observe a very sharp crossover from near perfect fidelity to a fidelity of zero, for a wave-packet with a velocity of $v_b$ i.e. $F\!\approx\! 0$ for $t_f\!<\!\frac{d}{v_b}$ while $F\!\approx\!1$ for $t_f\!>\!\frac{d}{v_b}$. For $\omega_0\!=\!0.5 J/\hbar$ shown in Fig. \ref{fig_sweepDT}(a) and (b) we numerically find $v_b\! \approx\!0.95 \Delta x \, J /\hbar$ which is slower than the maximum group velocity delineated by the red dotted lines and therefore also slower than the Lieb-Robinson velocity, shown by the white dot-dashed lines. For different trapping frequencies we find that $v_b$ behaves roughly linearly with $\omega_0$ with larger trapping frequencies (corresponding to tighter confined magnons) allowing for faster transport, as seen by comparing Figs.~\ref{fig_sweepDT}(a,b) with Figs.~\ref{fig_sweepDT}(c,d). While the limit velocity, $v_b$, varies for different protocols, $X_0$, the ramp considered has the lowest values of $v_b$ of the  schemes tested. Initial simulations suggest that reducing the frequency during the process results in a better fidelity and lower values of $v_b$, however a detailed study of this is left for another work.

\begin{figure}[t]
\begin{center}
 \includegraphics[width=0.9\linewidth]{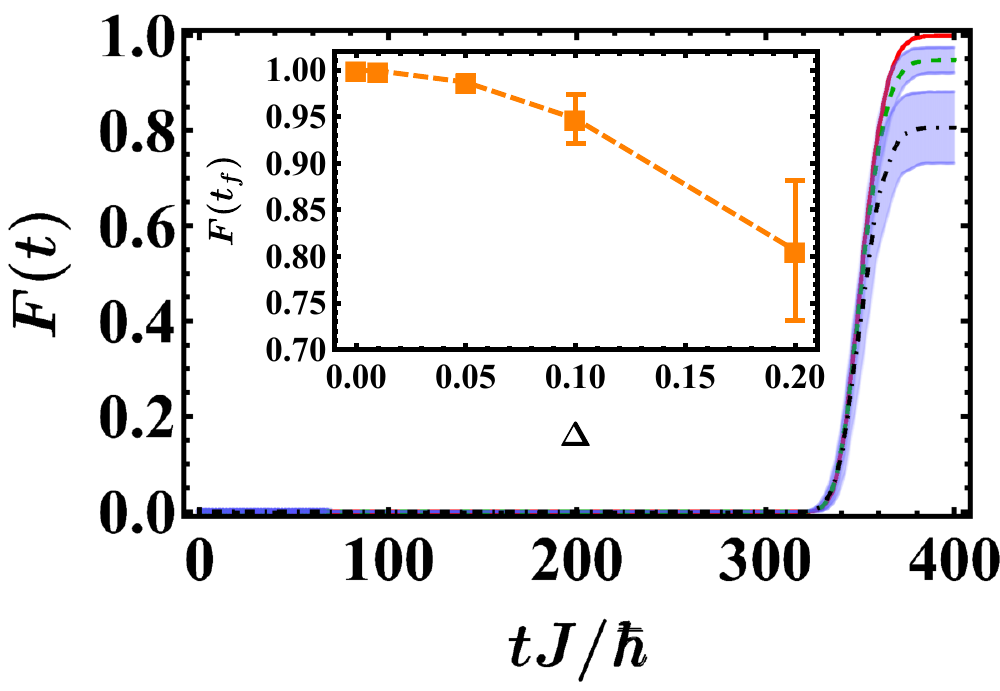} 
\end{center}
\caption{Fidelity as a function of time for nonadiabatic scheme \eqref{statraj} in the presence of disorder. Fidelity averaged over $1000$ realisations for noise amplitude $\Delta=0.01$ (red, solid line), $\Delta=0.1$ (green, dashed line) and $\Delta=0.2$ (black, dot-dashed line) with standard deviation error (blue, shaded region). Inset: Final fidelity as a function of noise amplitude $\Delta$ with standard deviation error bars. Parameters: $N=251$, $t_f=400 \, \hbar J^{-1}$, $\omega_0=0.5 \, J/\hbar$, $x_A = 50 \, \Delta x$ and $d=150 \, \Delta x$.
\label{fig_disorder}}
\end{figure}
\subsection{Effects of Disorder}
\label{disorder}
We finally investigate the robustness of the non-adiabatic protocol when there is disorder present in the system. We assume the total Hamiltonian is given by
\begin{equation}
H(t)= - \frac{1}{2} \sum_{n=1}^{N-1} J_n \vec{\sigma}_{n+1}\cdot \vec{\sigma}_n+ \sum_{n=1}^{N} B_n(t) \sigma_n^z,
\end{equation}
where $J_n\!=\!J(1+\epsilon_n)$~\cite{Gong2008,Ahmed2015}, $\epsilon_n$ is a random variable uniformly distributed over $[-\Delta,\Delta]$ and $\Delta$ characterises the amplitude of the noise. The fidelity of achieving the final target state in the presence of disorder, using the non-adiabatic protocol, is shown in Fig.~\ref{fig_disorder} after averaging over 1000 realisations of the disorder parameter and for $t_f\!=\! 400\hbar J^{-1}$. For comparison we show that the noise free case achieves a fidelity $F\!>\!0.999$ which subsequently decreases approximately quadratically with increasing noise amplitude. Nevertheless we find that for moderate amplitudes,  $\Delta\! \lesssim \!0.05$, the average fidelity remains greater than $0.98$ indicating that the protocol is robust to imperfect implementation. 
For short operation times $t_f$, disorder has a larger effect leading to lower target state fidelities, becoming particularly sensitive to these fluctuations near the limit $d/v_b$.

\section{Conclusion \label{con}}
In this work we have proposed a stable protocol to achieve fast high fidelity magnon transport in a one-dimensional spin chain. This was achieved by adapting analytic techniques for perfect control of a harmonically trapped particle to suit the discrete spin-chain. We exploited an approximate mapping between Hamiltonians for the spin-system and the trapped single particle. This allowed us employ the known Lewis-Risenfeld invariants to achieve near perfect magnon transport. Due to the approximate nature of the mapping, we numerically explored its performance in comparison to other known adiabatic schemes, establishing that our control protocol allows for significantly faster and more stable transport. We also showed that the protocol is robust to disorder. Furthermore, a natural speed limit emerged which was compared to other known bounds for transport in spin-systems, in particular the Lieb-Robinson bound and the maximal group velocity. While necessarily slower than these limits, our protocol nevertheless behaved comparably with these bounds. 

 There are several possibilities to extend these results. Errors from imperfections in the magnetic field such as anharmonicities ~\cite{Zhang2015}, stochastic fluctuations ~\cite{Lu2014} or incorrect trapping frequencies ~\cite{Guery2014} could be compensated for. Alternative settings could also be considered such as the transportation of two-magnon states~\cite{Longo2013} or using electric fields to modulate the inter-spin coupling~\cite{AhmedThesis}. Finally we note that spin chains with next nearest neighbour or long range interactions~\cite{Tony3} could be accounted for by using higher order finite difference approximations for the kinetic energy term in Eq. \eqref{singlep}.

Our results show that coherent control of complex many-body quantum systems can be achieved through a hybrid approach involving approximations and known analytic control techniques. Indeed, a common criticism of shortcuts to adiabaticity lies in their range of applicability. The fact that one must analytically solve the model severely limits the range of problems to which these techniques can be directly applied. Our work opens the possibility to significantly extend this range by establishing that, even if only approximately valid, the information learned from analytical control schemes can be invaluable for finding useful controls for many-body systems.
 
\section*{Acknowledgements}
 This work is supported by the Science Foundation Ireland Starting Investigator Research Grant “SpeedDemon” No. 18/SIRG/5508.

\bibliography{magnon}
\end{document}